\documentclass[10pt,conference]{IEEEtran}
\IEEEoverridecommandlockouts
\usepackage[numbers]{natbib}
\usepackage{pifont}
\usepackage{amsmath,amssymb,amsfonts}
\usepackage[table]{xcolor}
\usepackage{graphicx}
\usepackage{url}
\usepackage{multirow}
\usepackage{verbatim}
\usepackage{xurl}
\usepackage{dblfloatfix}    

\AtBeginDocument{%
  \providecommand\BibTeX{{%
    \normalfont B\kern-0.5em{\scshape i\kern-0.25em b}\kern-0.8em\TeX}}}

\newif\ifdraft
\draftfalse
\def\boldification #1 {\ifdraft\textbf{#1\newline\indent}\else\relax\fi}

\begin{document}

\title{Please Don't Go - A Comprehensive Approach to Increase Women's Participation in Open Source Software}

\author{\IEEEauthorblockN{Bianca Trinkenreich}
\IEEEauthorblockA{\textit{Northern of Arizona University} \\
\textit{Flagstaff, AZ, USA} \\
bt473@nau.edu}
}

\maketitle

\thispagestyle{plain}
\pagestyle{plain}

\begin{abstract}

Women represent less than 24\% of employees in the software development industry and experience various types of prejudice and bias. Despite various efforts to increase diversity and multi-gendered participation, women are even more underrepresented in Open Source Software (OSS) projects. In my PhD, I investigate the following question: How can OSS communities increase women's participation in their projects? I will identify different OSS career pathways and develop a holistic view of women's motivations to join or leave OSS, as well as their definitions of success. Based on this empirical investigation, I will work together with the Linux Foundation to design attraction and retention strategies focused on women. Before and after implementing the strategies, I will conduct empirical studies to evaluate the state of the practice and understand the implications of the strategies.

\end{abstract}

\begin{IEEEkeywords}
open source software, women, gender, diversity, participation, success, career
\end{IEEEkeywords}

\section{Problem and Research Statement}


Open Source Software (OSS) development is a collaborative endeavor in which expert developers distributed around the globe create software solutions~\cite{oreg2008exploring, forte2013defining}. Many OSS projects count on a community of volunteers to succeed, and such a community needs newcomers for their sustainability and growth. 
The lack of gender diversity in OSS projects has gained increasing attention from practitioners and researchers. 


Diversity in software development teams can take many different forms, including gender, experience, culture, and technical knowledge. Some teams are more diverse in one attribute and less in others~\cite{vasilescu2015gender}. 
Gender diversity positively affects productivity by bringing together different perspectives; improving outcomes~\cite{vasilescu2015gender}, innovation, and problem-solving capacity; and fostering a healthier work environment~\cite{earley2000creating}. A diverse development team is more likely to properly comprehend users’ needs, contributing to a better alignment between the delivered software and its intended customers~\cite{muller1993participatory}.

\boldification{***Women numbers in OSS are low***}
Although organizations are taking actions to increase gender diversity, the percentage of women in OSS projects are in average lower than 10\%. Only 7.5\% of the contributions to public code from the last 50 years were authored by women~\cite{zacchiroli2020gender}. Women represent only 5.2\% of the contributors to the Apache Software Foundation~\cite{asf2016survey}, 9.9\% in Linux kernel~\cite{bitergia2016survey}, and 10\% of OpenStack contributors~\cite{izquierdo2018openstack}, three of the largest and most well-known OSS communities. Indeed, women represent only 9\% of GitHub users~\cite{vasilescu2015data}. 

Considering the benefits of having a more gender-diverse team, researchers are also increasingly focusing on understanding the low representation of women in OSS. 

\boldification{***Gender bias is one of the barriers that cause low numbers of women ***}

Research suggests that gender bias and sexist behavior pervade OSS~\cite{nafus2012patches,terrell2017gender}. Women feel frustrated when they are the only woman on a development team or when their input is under-valued or ignored, even on topics in which they have expert knowledge~\cite{vasilescu2015perceptions}. Within OSS projects, the notion of meritocracy reigns, following the logic that quality speaks for itself and will be rewarded~\cite{feller2000framework}. Continually finding themselves on the bottom rung, it is no surprise that many women report experiencing ``imposter syndrome"~\cite{vasilescu2015gender}. Gender biases can represent a a persistent barrier for women to join OSS~\cite{mendez2018open}.



Strategies suggested by previous work to attract and retain women include issuing code of conduct statements~\cite{tourani2017code,lee2019floss,fossatti2020gender}; adopting feminist and social justice principles~\cite{fossatti2020gender}; promoting women to leadership positions~\cite{bosu2019diversity};
providing spaces for women to build their leadership capacity and engage in developing the community with
norms and values consistent with their own~\cite{qiu2010joining}, 
focusing on the first social experiences through programs such as mentorships~\cite{kuechler2012gender}; and reforming the systemic gender-bias and providing inclusive tools and infrastructure~\cite{mendez2018open}. Strategies discouraged by the literature include setting quotas for women, since merely increasing the proportion of women can lead to the flattening of the slope of the relationship between behavioral femaleness and outcomes and activate questionable stereotypes~\cite{vedres2019gendered}.


Consistent with the general finding that women's participation in OSS remains very low, survey and anecdotal evidence have indicated that attracting and retaining women as contributors in OSS projects has been particularly challenging~\cite{ghosh2002free,weiss2005panel,wurzelova2019characterizing}.


\section{Research Goals and Hypothesis}



\begin{figure}[hbt]
     \centering
     \includegraphics[width=0.48\textwidth]{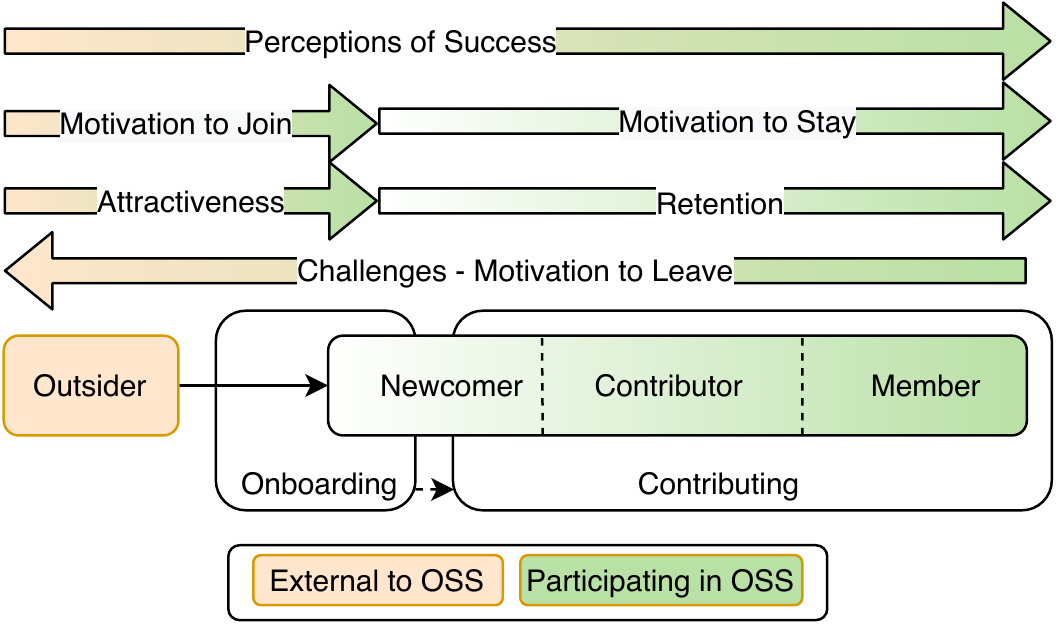}
     \caption{Contributors joining model presenting the stages and forces that act during the joining process}
     \label{fig:stages_forces}
 \end{figure}

I will investigate women's participation in OSS, including their motivations to join, what attracts them to OSS, the pathways they follow as they progress on a project (or to achieve their perceptions of success), the challenges they face and their motivations to stay, take breaks, or leave, and what attracts or repels them from OSS. Ultimately, the overall goal of this project is to help OSS projects devise strategies to attract and retain women, while helping these women contributors to attain their own goals. For this purpose, I will create strategies to 1. attract and retain women based on the forces compelling them to join and stay in OSS, and 2. diminish the forces compelling them to drop out of OSS, as depicted in Fig. \ref{fig:stages_forces}.

The process of joining OSS projects can be understood into two stages:  \textsc{Onboarding} and \textsc{Contributing}~\cite{steinmacher2014attracting}. Explaining the process in different stages helps to visualize the different forces that compel contributors towards staying with or leaving the project~\cite{steinmacher2013newcomers}. 

Motivations drive both the \textsc{onboarding} and \textsc{contributing} stages of joining OSS. An individual can have one or a set of \textsc{motivation(s) to join}, and shift to (a) different \textsc{motivation(s) to stay} (or not)~\cite{ghosh2002free}. Shifts in motivation might occur due to changes in the OSS landscape, or they might reflect the journey an individual makes and their growth since first joining~\cite{von2012carrots,steinmacher2014attracting}. 

Individuals do not only behave to achieve immediate rewards; they might also act to reach or maintain a consistency of action points beyond the attainment of specific and immediate goals~\cite{von2012carrots}. In addition to motivations, self-perceptions of success affect choices they make for personal and professional lives, including educational options and decisions of where to work~\cite{dyke2006we}. Aligned with this, I argue that an OSS community or organization can better attract and retain women contributors when they consider both their motivations and the multitude of factors that underpin what means success to them. 


I will identify the challenges that women report when contributing to OSS, and what would make them take breaks or even leave altogether. I will collect their advice for other women and their suggestions about possible actions to increase inclusivity, and also use as input for the strategies to attract and retain them.


In this dissertation, I tackle the research question \textbf{How can OSS communities increase women’s  participation in OSS projects?}. To guide my exploration of an answer to the research question, I defined specific questions: 
\begin{itemize}
    \item \textit{RQ.1 What strategies can be employed to attract and retain more women contributors to OSS projects?}
    \item\textit{RQ.2. How can these strategies help to increase the percentage of women in OSS projects?}
\end{itemize}



\section{Expected Contributions}

The theoretical contribution of this dissertation is multi-fold, including identification of: the pathways that can be followed by OSS contributors (from any gender); women's motivations to join (or not), stay, take breaks, and leave OSS projects; women contributors' multi-faceted definitions of success; contributions to the state-of-the-art, including the current challenges women face, women's advice for other women, and suggestions to make OSS projects more inclusive.

The practical contributions include guidelines for women seeking a career in OSS, showing the different roles, activities, backgrounds, and necessary skills. Women can use the guidelines to develop a training plan, and learn and improve the skills necessary for their preferred pathway. I will also provide actionable mechanisms for OSS projects to encourage women to join and keep contributing. The strategies will be implemented in a large OSS project that seeks to increase women's participation (Linux Kernel) and evaluated in practice, as I explain in Section \ref{stage3}.


To the best of my knowledge, there is no work that provides strategies to increase women's participation in OSS based on what I call ``women's desires and beliefs about OSS;'' that is, their motivations to join, stay, or leave, their perceptions of success, and the challenges they encounter. Also there is no work that offers a guideline of career pathways and roles in OSS that highlights for women the different ways they can contribute, be successful, and achieve their goals.


\section{Research Methodology}

\begin{figure*}[hbt]
     \centering
     \includegraphics[width=1\textwidth]{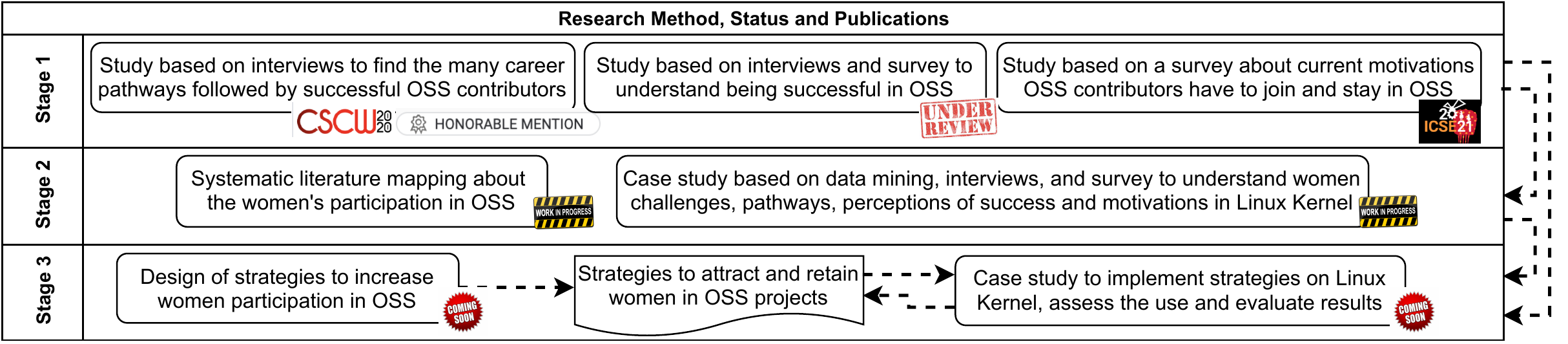}
     \caption{Research Design}
     \label{fig:research_design}
 \end{figure*}

My research has three stages and adopts mixed methods to accomplish its goal, as depicted in Figure \ref{fig:research_design}.

To help answer \textit{RQ.1. What strategies can be employed to attract and retain more women as contributors to OSS projects?} of this dissertation, I've executed Stage 1 and planned Stage 2 to explore the career pathways, goals, definitions of success, and motivations that influence a contributor's decision to join, stay, take breaks, or leave an OSS project. The studies include all genders contributors as I use segmented analysis to compare findings between genders.


\subsection{Stage 1 - Explore career pathways, motivations, and definitions of success}
\label{stage1}

I started this project with a \textbf{study based on interviews to find the different pathways followed by successful OSS contributors~\cite{trinkenreich2020pathways}}. We interviewed 17 participants (12 of them identified their gender as women) that were invited speakers at OSCON (Open Source Software Conference). I have qualitatively analyzed the results to identify career paths, how they joined, which roles and activities they perform, and how they arrived at their current position in OSS. Understanding that participation in OSS projects includes more than writing code~\cite{izquierdo2018openstack}, we found that people can build a career in OSS through different roles and activities, and with different backgrounds, including those not related to writing software. These activities, while crucial to the survival of the OSS community, are currently performed by hidden or largely unacknowledged figures in the community. They (or their contributions) are not visible when navigating the project repository data. As a result they are not seen as the central figures of the project. This not only is demotivating, because of the lack of recognition, but it can also discourage individuals who lack a Computer Science background or interest in coding-related activities.


Next, I conducted a second \textbf{study based on interviews and a survey to understand the multi-faceted definition of success in OSS}. 
Success in OSS encompasses more than code contributions alone; as evident in one participant's OSS journey, who shared in our first study that success in OSS involves, who said it involves \textit{``contributing more than code, [including] contributing documentation, processes \ldots the governance of the project''} (P7). However, currently there is a misconception that success in OSS is only achieved through activities related to source code~\cite{lakhani2003hackers, fitzgerald2006transformation, robles2019twenty, Steinmacher.Teenager:2017}. 

The disproportionate emphasis on code can make other types of contributions seem less valuable. This may specifically disadvantage women, given that the majority of code authors are men, and social ties may influence women's inclusion in technical activities~\cite{nafus2012patches,qiu2019going} resulting in women's lower engagement~\cite{Qiu.ea.2019}. Affirming this, the majority (144 out of 165) of men who answered our survey were coders. OSS contributors, however, comprise a heterogeneous group, with diverse talents, skills, career goals, and motivations~\cite{harsworking, hertel2003motivation, von2012carrots, ghosh2005understanding}. Also from the first study, we found that some contributors perform a variety of non-code related activities (e.g., advocacy, technical writing, translation, project management)~\cite{trinkenreich2020pathways} and follow different pathways than the celebrated ``onion model''~\cite{nafus2012patches, trainer2015personal, trinkenreich2020pathways}. 

Given the fact that OSS communities comprise more players than simply ``code warriors,'' the community's definition of success ought to broaden beyond the quantity of code one produces. The way we define success has a remarkable impact on the choices we make in our personal and professional lives. Without such a broadened understanding, how would it possible support diverse individuals whose background, career goals, and pathways do not fit the typical onion model career mold? I have investigated the different career pathways, goals, and the self-definitions of  success through interviews with 27 OSS contributors who are recognized as successful in their communities, and a follow-up open survey with 193 OSS contributors. This study provides nuanced definitions of success perceptions in OSS, and show that OSS contributors have a broader perspective on success than the narrow focus on code-related activities---which is  better supported by current tools and practices. Our results include 26 categories of definitions of success. A segmented analysis by gender showed that women consider recognition more than men as part of their definition for success. The literature shows that men relate success to tangible and objective outcomes, but, contrary to the research in other domains~\cite{dyke2006we, cho2017south, porter2019physics}, definitions of success that are considered subjective were also cited by men. This study was also turned into a paper that is currently under peer review.

Following the first two studies, I conducted a third \textbf{study based on a survey about the current motivations that drive OSS contributors to join and stay in OSS}. Our field’s understanding of what motivates people to contribute to OSS is still fundamentally grounded in studies from the early  2000s, and much has changed since the early days of OSS~\cite{steinmacher2017free}. OSS today enjoys a place of distinction in producing key technologies and providing learning; from the first study we could see that OSS also offers different career opportunities. With such drastic changes to the status of OSS, 
we considered it likely that what motivates people to join OSS also has evolved since the early days. 
It is time to revisit the fundamental question of what drives people to contribute to OSS today. Shifts in motivation occur not only due to  changes to the OSS landscape, but also in reflection of the journey an individual makes and their growth since they first joined~\cite{steinmacher2014attracting}. Currently, there is a lack an understanding of the differences in motivation for the early joiners compared to those who are well-entrenched in OSS.

Aiming to support both the attraction of new members and the retention of existing contributors, we ran this study to understand the current motivations of OSS contributors, how they shifted from OSS's infancy~\cite{lakhani2003hackers, harsworking} in response to the changing landscape, and how motivation changes after  members join OSS. Through a survey with 242 OSS contributors, the results showed that intrinsic and internalized motivations currently explain what drives most contributors today. On the extrinsic end, \emph{career} is relevant to many contributors, contrary to \emph{pay}, which only explains why less than one-third of the respondents contributing to OSS. Contributing to OSS often transforms extrinsic motivations when joining to intrinsic ones to stay. Whereas ideology, own-use, or education-related programs can be an impetus to join OSS, individuals stay for intrinsic reasons (fun, altruism, reputation, and kinship). Results also showed that some motivations to contribute to OSS have endured since the 2000s: learning, fun, knowledge sharing, and a belief that source code should be open---all core tenets of OSS. Others have seen a marked difference. Social aspects (e.g., altruism, kinship, and reputation) have increased in the ranking, whereas participating in OSS to ``scratch one's own itch'' has dropped. Very few women from the respondents joined OSS because of traditional motivations, but continued for \emph{reciprocity}, which was one of the lowest factors for men. This study was accepted and will be presented at ICSE 2021 in the Research Track.

From the studies performed in Stage 1, I observed that although motivation and success perceptions are interrelated and complement each other, they can play different roles. As an example, one of our participants reported being motivated to join by “reputation”, but perceived success as “getting paid to contribute”. Individuals with diverse backgrounds and understandings of success may need different engagement strategies~\cite{trinkenreich2020pathways}. By understanding that success is polyvalent in OSS, communities can leverage different definitions of success to support the growth of diverse individuals. For contributors who consider success as "Having contacts in several different communities", communities can promote meetups to help increase social capital.

\subsection{Stage 2 - Understanding women's participation in OSS}
\label{stage2}

I am currently analyzing the results of a systematic literature review to identify studies about women's participation in OSS projects, including what types of activities women perform in OSS projects, their motivations to join, stay, not join, and leave, challenges they face, and strategies to increase women's participation in OSS projects. The search string was be executed in ACM, Scopus, Compendex and IEEE.

In parallel with the literature review, I am running a \textbf{case study based on survey and interviews to understand women's challenges, career pathways, perceptions of success, and motivations in the Linux Kernel}. I will start with an on-line survey with all the current contributors Linux Kernel. Besides demographics, the survey's questionnaire will include open and closed questions about motivations, success, challenges, advice for their women peers, and suggestions to increase  women's participation and suggestions for metrics to evaluate the sense of virtual community~\cite{blanchard2007developing,abfalter2012sense}. After qualitative and quantitative analysis of the data, I will corroborate the findings through semi-structured interviews~\cite{mihas2019learn}. For the interviews I will first identify survey respondents who had agreed to follow up conversations. From this set of respondents, I will randomly select a set of interview participants to balance the demographics distribution and include different genders, as well those who had never taken breaks and those who contributed on a regular basis, then took a break, and came back. In addition, I will mine Linux Kernel repositories to identify contributors who contributed on a regular basis for a certain time and left, and try to recruit them for an interview with the Linux Kernel community managers.
Following this, I will interview and qualitatively analyze data using coding procedures and compare different genders on: their motivations to join, stay, take breaks, and leave; the categories of challenges they face; how they evolve and the pathways they take; how they perceive success; and their suggestions for improving inclusivity. Added to this data, I will collect information about the results of the actions this project had already taken to increase inclusivity. 

\subsection{Stage 3 - Strategies to Increase Women in OSS}
\label{stage3}

In this Stage I will design actionable strategies to increase women's participation. The interventions will focus on increasing their sense of belonging and stickiness with the project. The community managers of Linux Kernel will evaluate the feasibility of the strategies. The data collected in both Stages 1 and 2 will be used as design input.

In order to answer the \textit{RQ.2. How can the strategies help increase the rates of women in OSS projects?}, and aligned with the Plan-Do-Check-Act (PDCA) approach~\cite{shewhart1986statistical,deming2018out}, during the execution of the strategies I will interact and collect feedback from the contributors and the community managers to evaluate, learn, and improve the strategies. Additionally, I will run post-study debriefing interview sessions with women who participated in the strategies to collect their impressions, and the positive and negative points to be used as lessons learned. I will track of the number of women before and after the strategies were implemented to evaluate whether the strategies helped to increase the number of women contributors.

\section{Conclusion}

This proposal focuses on increasing women's participation in OSS projects. More specifically, I take an holistic approach to women's motivations to join (or not), stay, take breaks, or leave OSS, together with their definitions of success, to design attraction and retention strategies focused on women's aspirations within OSS. Through preliminary findings, I show the many different career pathways that women can follow find success in OSS, how their motivations to start and to stay can shift, and that their definition of success is multi-faceted and nuanced. Success can include both objective perspectives (e.g., monetary rewards, amount of contribution) as well as subjective perceptions (e.g., recognition in the community, satisfaction).
Although contributors' goals may end up at the same destination, the routes they take to arrive there can be many and divergent, rooted in their diverse motivations and perceptions of success.
\bibliographystyle{IEEEtran}
\bibliography{paper}

\begin{thebibliography}{10}
\providecommand{\url}[1]{#1}
\csname url@samestyle\endcsname
\providecommand{\newblock}{\relax}
\providecommand{\bibinfo}[2]{#2}
\providecommand{\BIBentrySTDinterwordspacing}{\spaceskip=0pt\relax}
\providecommand{\BIBentryALTinterwordstretchfactor}{4}
\providecommand{\BIBentryALTinterwordspacing}{\spaceskip=\fontdimen2\font plus
\BIBentryALTinterwordstretchfactor\fontdimen3\font minus
  \fontdimen4\font\relax}
\providecommand{\BIBforeignlanguage}[2]{{%
\expandafter\ifx\csname l@#1\endcsname\relax
\typeout{** WARNING: IEEEtran.bst: No hyphenation pattern has been}%
\typeout{** loaded for the language `#1'. Using the pattern for}%
\typeout{** the default language instead.}%
\else
\language=\csname l@#1\endcsname
\fi
#2}}
\providecommand{\BIBdecl}{\relax}
\BIBdecl

\bibitem{oreg2008exploring}
S.~Oreg and O.~Nov, ``Exploring motivations for contributing to open source
  initiatives: The roles of contribution context and personal values,''
  \emph{Computers in human behavior}, vol.~24, no.~5, pp. 2055--2073, 2008.

\bibitem{forte2013defining}
A.~Forte and C.~Lampe, ``Defining, understanding, and supporting open
  collaboration: Lessons from the literature,'' \emph{American Behavioral
  Scientist}, vol.~57, no.~5, pp. 535--547, 2013.

\bibitem{vasilescu2015gender}
B.~Vasilescu, D.~Posnett, B.~Ray, M.~G. van~den Brand, A.~Serebrenik,
  P.~Devanbu, and V.~Filkov, ``Gender and tenure diversity in github teams,''
  in \emph{Proceedings of the 33rd annual ACM conference on human factors in
  computing systems}, 2015, pp. 3789--3798.

\bibitem{earley2000creating}
C.~P. Earley and E.~Mosakowski, ``Creating hybrid team cultures: An empirical
  test of transnational team functioning,'' \emph{Academy of Management
  journal}, vol.~43, no.~1, pp. 26--49, 2000.

\bibitem{muller1993participatory}
M.~J. Muller and S.~Kuhn, ``Participatory design,'' \emph{Communications of the
  ACM}, vol.~36, no.~6, pp. 24--28, 1993.

\bibitem{zacchiroli2020gender}
S.~Zacchiroli, ``Gender differences in public code contributions: a 50-year
  perspective,'' \emph{IEEE Software}, 2020.

\bibitem{asf2016survey}
F.~Sharan, ``Asf committer diversity survey. accessed: 2020-10-16,''
  \url{https://cwiki.apache.org/confluence/display/COMDEV/ASF+Committer+Diversity+Survey+-+2016},
  2016.

\bibitem{bitergia2016survey}
Bitergia, ``Gender-diversity analysis of the linux kernel technical
  contributions. accessed: 2020-10-16,''
  \url{https://blog.bitergia.com/2016/10/11/gender-diversity-analysis-of-the-linux-kernel-technical-contributions},
  2016.

\bibitem{izquierdo2018openstack}
D.~Izquierdo, N.~Huesman, A.~Serebrenik, and G.~Robles, ``Openstack gender
  diversity report,'' \emph{IEEE Software}, vol.~36, no.~1, pp. 28--33, 2018.

\bibitem{vasilescu2015data}
B.~Vasilescu, A.~Serebrenik, and V.~Filkov, ``A data set for social diversity
  studies of github teams,'' in \emph{2015 IEEE/ACM 12th working conference on
  mining software repositories}.\hskip 1em plus 0.5em minus 0.4em\relax IEEE,
  2015, pp. 514--517.

\bibitem{nafus2012patches}
D.~Nafus, ``‘patches don’t have gender’: What is not open in open source
  software,'' \emph{New Media \& Society}, vol.~14, no.~4, pp. 669--683, 2012.

\bibitem{terrell2017gender}
J.~Terrell, A.~Kofink, J.~Middleton, C.~Rainear, E.~Murphy-Hill, C.~Parnin, and
  J.~Stallings, ``Gender differences and bias in open source: Pull request
  acceptance of women versus men,'' \emph{PeerJ Computer Science}, vol.~3, p.
  e111, 2017.

\bibitem{vasilescu2015perceptions}
B.~Vasilescu, V.~Filkov, and A.~Serebrenik, ``Perceptions of diversity on git
  hub: A user survey,'' in \emph{2015 IEEE/ACM 8th International Workshop on
  Cooperative and Human Aspects of Software Engineering}.\hskip 1em plus 0.5em
  minus 0.4em\relax IEEE, 2015, pp. 50--56.

\bibitem{feller2000framework}
J.~Feller and B.~Fitzgerald, ``A framework analysis of the open source software
  development paradigm,'' in \emph{ICIS 2000 proceedings of the twenty first
  international conference on information systems}.\hskip 1em plus 0.5em minus
  0.4em\relax Association for Information Systems (AIS), 2000, pp. 58--69.

\bibitem{mendez2018open}
C.~Mendez, H.~S. Padala, Z.~Steine-Hanson, C.~Hilderbrand, A.~Horvath, C.~Hill,
  L.~Simpson, N.~Patil, A.~Sarma, and M.~Burnett, ``Open source barriers to
  entry, revisited: A sociotechnical perspective,'' in \emph{Proceedings of the
  40th International Conference on Software Engineering}, 2018, pp. 1004--1015.

\bibitem{tourani2017code}
P.~Tourani, B.~Adams, and A.~Serebrenik, ``Code of conduct in open source
  projects,'' in \emph{2017 IEEE 24th international conference on software
  analysis, evolution and reengineering (SANER)}.\hskip 1em plus 0.5em minus
  0.4em\relax IEEE, 2017, pp. 24--33.

\bibitem{lee2019floss}
A.~Lee and J.~C. Carver, ``Floss participants' perceptions about gender and
  inclusiveness: a survey,'' in \emph{2019 IEEE/ACM 41st International
  Conference on Software Engineering (ICSE)}.\hskip 1em plus 0.5em minus
  0.4em\relax IEEE, 2019, pp. 677--687.

\bibitem{fossatti2020gender}
M.~Fossatti, ``Gender, diversity, and inclusion in open source communities,''
  \emph{XRDS: Crossroads, The ACM Magazine for Students}, vol.~26, no.~4, pp.
  46--48, 2020.

\bibitem{bosu2019diversity}
A.~Bosu and K.~Z. Sultana, ``Diversity and inclusion in open source software
  (oss) projects: Where do we stand?'' in \emph{2019 ACM/IEEE International
  Symposium on Empirical Software Engineering and Measurement (ESEM)}.\hskip
  1em plus 0.5em minus 0.4em\relax IEEE, 2019, pp. 1--11.

\bibitem{qiu2010joining}
Y.~Qiu, K.~J. Stewart, and K.~M. Bartol, ``Joining and socialization in open
  source women’s groups: an exploratory study of kde-women,'' in \emph{IFIP
  International Conference on Open Source Systems}.\hskip 1em plus 0.5em minus
  0.4em\relax Springer, 2010, pp. 239--251.

\bibitem{kuechler2012gender}
V.~Kuechler, C.~Gilbertson, and C.~Jensen, ``Gender differences in early free
  and open source software joining process,'' in \emph{IFIP International
  Conference on Open Source Systems}.\hskip 1em plus 0.5em minus 0.4em\relax
  Springer, 2012, pp. 78--93.

\bibitem{vedres2019gendered}
B.~Vedres and O.~Vasarhelyi, ``Gendered behavior as a disadvantage in open
  source software development,'' \emph{EPJ Data Science}, vol.~8, no.~1, p.~25,
  2019.

\bibitem{ghosh2002free}
R.~Ghosh, R.~Glott, B.~Krieger, and G.~Robles, ``Free/libre and open source
  software: Survey and study,'' 2002.

\bibitem{weiss2005panel}
T.~Weiss, ``Panel: Open-source needs more women developers,''
  \emph{Computerworld}, 2005.

\bibitem{wurzelova2019characterizing}
P.~Wurzelov{\'a}, F.~Palomba, and A.~Bacchelli, ``Characterizing women (not)
  contributing to open-source,'' in \emph{2019 IEEE/ACM 2nd International
  Workshop on Gender Equality in Software Engineering (GE)}.\hskip 1em plus
  0.5em minus 0.4em\relax IEEE, 2019, pp. 5--8.

\bibitem{steinmacher2014attracting}
I.~Steinmacher, M.~A. Gerosa, and D.~Redmiles, ``Attracting, onboarding, and
  retaining newcomer developers in open source software projects,'' in
  \emph{Workshop on GSD in a CSCW Perspective}, 2014.

\bibitem{steinmacher2013newcomers}
I.~Steinmacher, I.~Wiese, A.~P. Chaves, and M.~A. Gerosa, ``Why do newcomers
  abandon open source software projects?'' in \emph{2013 6th International
  Workshop on Cooperative and Human Aspects of Software Engineering
  (CHASE)}.\hskip 1em plus 0.5em minus 0.4em\relax IEEE, 2013, pp. 25--32.

\bibitem{von2012carrots}
G.~Von~Krogh, S.~Haefliger, S.~Spaeth, and M.~W. Wallin, ``Carrots and
  rainbows: Motivation and social practice in open source software
  development,'' \emph{MIS quarterly}, pp. 649--676, 2012.

\bibitem{dyke2006we}
L.~S. Dyke and S.~A. Murphy, ``How we define success: A qualitative study of
  what matters most to women and men,'' \emph{Sex Roles}, vol.~55, no. 5-6, pp.
  357--371, 2006.

\bibitem{trinkenreich2020pathways}
B.~Trinkenreich, M.~Guizani, I.~Wiese, A.~Sarma, and I.~Steinmacher, ``Hidden
  figures: Roles and pathways of successful oss contributors,'' \emph{Computer
  Supported Cooperative Work (CSCW)}, vol.~4, no. 180, 2020.

\bibitem{lakhani2003hackers}
K.~R. Lakhani and R.~G. Wolf, ``Why hackers do what they do: Understanding
  motivation and effort in free/open source software projects,'' 2003.

\bibitem{fitzgerald2006transformation}
B.~Fitzgerald, ``The transformation of open source software,'' \emph{MIS
  quarterly}, vol.~30, no.~3, pp. 587--598, 2006.

\bibitem{robles2019twenty}
G.~Robles, I.~Steinmacher, P.~Adams, and C.~Treude, ``Twenty years of open
  source software: From skepticism to mainstream,'' \emph{IEEE Software},
  vol.~36, no.~6, pp. 12--15, 2019.

\bibitem{Steinmacher.Teenager:2017}
\BIBentryALTinterwordspacing
I.~Steinmacher, G.~Robles, B.~Fitzgerald, and A.~Wasserman, ``Free and open
  source software development: the end of the teenage years,'' \emph{Journal of
  Internet Services and Applications}, vol.~8, no.~1, p.~17, Dec 2017.
  [Online]. Available: \url{https://doi.org/10.1186/s13174-017-0069-9}
\BIBentrySTDinterwordspacing

\bibitem{qiu2019going}
H.~S. Qiu, A.~Nolte, A.~Brown, A.~Serebrenik, and B.~Vasilescu, ``Going farther
  together: The impact of social capital on sustained participation in open
  source,'' in \emph{2019 IEEE/ACM 41st International Conference on Software
  Engineering (ICSE)}.\hskip 1em plus 0.5em minus 0.4em\relax IEEE, 2019, pp.
  688--699.

\bibitem{Qiu.ea.2019}
\BIBentryALTinterwordspacing
------, ``Going farther together: The impact of social capital on sustained
  participation in open source,'' in \emph{Proceedings of the 41st
  International Conference on Software Engineering}, ser. ICSE '19.\hskip 1em
  plus 0.5em minus 0.4em\relax Piscataway, NJ, USA: IEEE Press, 2019, pp.
  688--699. [Online]. Available: \url{https://doi.org/10.1109/ICSE.2019.00078}
\BIBentrySTDinterwordspacing

\bibitem{harsworking}
A.~Hars and S.~Ou, ``Working for free motivations of participating in open
  source software projects,'' in \emph{HICSS'04}, 2004, pp. 25--31.

\bibitem{hertel2003motivation}
G.~Hertel, S.~Niedner, and S.~Herrmann, ``Motivation of software developers in
  open source projects: an internet-based survey of contributors to the linux
  kernel,'' \emph{Research policy}, vol.~32, no.~7, pp. 1159--1177, 2003.

\bibitem{ghosh2005understanding}
R.~A. Ghosh, ``Understanding free software developers: Findings from the floss
  study,'' \emph{Perspectives on free and open source software}, vol.~28, pp.
  23--47, 2005.

\bibitem{trainer2015personal}
E.~H. Trainer, C.~Chaihirunkarn, A.~Kalyanasundaram, and J.~D. Herbsleb, ``From
  personal tool to community resource: What's the extra work and who will do
  it?'' in \emph{18th ACM Conference on Computer Supported Cooperative Work \&
  Social Computing}, 2015, pp. 417--430.

\bibitem{cho2017south}
Y.~Cho, J.~Park, S.~Jeoung, B.~Ju, J.~You, A.~Ju, C.~K. Park, H.~Y. Park
  \emph{et~al.}, ``How do south korean female executives’ definitions of
  career success differ from those of male executives?'' \emph{European Journal
  of Training and Development}, 2017.

\bibitem{porter2019physics}
A.~M. Porter, ``Physics phds ten years later: Success factors and barriers in
  career paths. results from the phd plus 10 study.'' \emph{AIP Statistical
  Research Center}, 2019.

\bibitem{steinmacher2017free}
I.~Steinmacher, G.~Robles, B.~Fitzgerald, and A.~Wasserman, ``Free and open
  source software development: the end of the teenage years,'' \emph{Journal of
  Internet Services and Applications}, vol.~8, no.~17, 2017.

\bibitem{blanchard2007developing}
A.~L. Blanchard, ``Developing a sense of virtual community measure,''
  \emph{CyberPsychology \& Behavior}, vol.~10, no.~6, pp. 827--830, 2007.

\bibitem{abfalter2012sense}
D.~Abfalter, M.~E. Zaglia, and J.~Mueller, ``Sense of virtual community: A
  follow up on its measurement,'' \emph{Computers in Human Behavior}, vol.~28,
  no.~2, pp. 400--404, 2012.

\bibitem{mihas2019learn}
P.~Mihas, \emph{Learn to Use an Exploratory Sequential Mixed Method Design for
  Instrument Development}.\hskip 1em plus 0.5em minus 0.4em\relax SAGE
  Publications, Limited, 2019.

\bibitem{shewhart1986statistical}
W.~A. Shewhart and W.~E. Deming, \emph{Statistical method from the viewpoint of
  quality control}.\hskip 1em plus 0.5em minus 0.4em\relax Courier Corporation,
  1986.

\bibitem{deming2018out}
W.~E. Deming, \emph{Out of the Crisis}.\hskip 1em plus 0.5em minus 0.4em\relax
  MIT press, 2018.

\end{thebibliography}

\end{document}